# Muon spin rotation study of magnetism and superconductivity in Ba(Fe$_{1-x}$Co$_x$)$_2$As$_2$ single crystals

C. Bernhard[1], C. N. Wang[1], L. Nuccio[1,2], L. Schulz[1*], O. Zaharko[3], J. Larsen[4], C. Aristizabal[2] M. Willis[2], A. J. Drew[2], G. D. Varma[5], T. Wolf[6], and Ch. Niedermayer[3]

[1]*University of Fribourg, Department of Physics and Fribourg Centre for Nanomaterials, Chemin du Musée 3, CH-1700 Fribourg, Switzerland*

[2]*Queen Mary University of London, School of Physics and Astronomy, Mile End Road, London E1 4NS, UK*

[3]*Laboratory for Neutron Scattering, Paul Scherrer Institut, CH-5232 Villigen, Switzerland*

[4]*Department of Physics, Technical University of Denmark, 2800 Lyngby, Denmark*

[5]*Department of Physics, Indian Institute of Technology Roorkee, Roorkee 247667, India*

[6]*Karlsruher Institut für Technologie, Institut für Festkörperphysik, D-76021 Karlsruhe, Germany*

**Abstract**

Using muon spin rotation ($\mu$SR) we investigated the magnetic and superconducting properties of a series of Ba(Fe$_{1-x}$Co$_x$)$_2$As$_2$ single crystals with $0 \leq x \leq 0.15$. Our study details how the antiferromagnetic order is suppressed upon Co substitution and how it coexists with superconductivity. In the non-superconducting samples at $0 < x < 0.04$ the antiferromagnetic order parameter is only moderately suppressed. With the onset of superconductivity this suppression becomes faster and it is most rapid between $x = 0.045$ and 0.05. As was previously demonstrated by $\mu$SR at $x = 0.055$ [P. Marsik *et al.*, Phys. Rev. Lett. **105**, 57001 (2010)], the strongly weakened antiferromagnetic order is still a bulk phenomenon that competes with superconductivity. The comparison with neutron diffraction data suggests that the antiferromagnetic order remains commensurate whereas the amplitude exhibits a spatial variation that is likely caused by the randomly distributed Co atoms. A different kind of magnetic order that was also previously identified [C. Bernhard *et al.*, New J. Phys. **11**, 055050 (2009)] occurs at $0.055 < x < 0.075$ where $T_c$ approaches the maximum value. The magnetic order develops here only in parts of the sample volume and it seems to cooperate with superconductivity since its onset temperature coincides with $T_c$. Even in the strongly overdoped regime at $x = 0.11$, where the static magnetic order has disappeared, we find that the low energy spin fluctuations are anomalously enhanced below $T_c$. These findings point toward a drastic change in the relationship between the magnetic and superconducting orders from a competitive one in the strongly underdoped regime to a constructive one in near optimally and overdoped samples.

1. **Introduction**

The discovery of high temperature superconductivity in the iron arsenides in 2008[1] with $T_c$ values as high as 55 K in ReFeAsO$_{1-x}$F$_x$ (Re = Sm, Nd and Gd)[2,3] has prompted intense experimental and theoretical efforts to explore their electronic properties. It turned out that similar to the cuprates, superconductivity emerges here in close proximity to an antiferromagnetic state. In the most commonly investigated system (Ba,Sr)Fe$_2$As$_2$, for which large single crystals are readily available, superconductivity can be likewise introduced by electron doping as in (Ba,Sr)(Fe$_{1-x}$Co$_x$)$_2$As$_2$,[4] by hole doping as in (Ba,Sr)$_{1-x}$K$_x$Fe$_2$As$_2$,[5,6] with external pressure[7,8] or with internal, chemical pressure as in (Ba,Sr)(Fe$_{1-x}$Ru$_x$)$_2$As$_2$.[9] It is commonly observed that the long-range AF order of the undoped parent compound (Ba,Sr)Fe$_2$As$_2$ becomes suppressed and superconductivity emerges even before the magnetic order has entirely disappeared.[10] The critical temperature of the superconducting state rises at first in the so-called underdoped regime where it coexists with a strongly weakened AF order. The highest critical temperature is obtained close to the critical point where the static magnetic order vanishes. Upon further doping or pressure, $T_c$ decreases again and finally disappears in the so-called overdoped regime. Since $T_c$ is maximal right at the point where static magnetism disappears and presumably low energy magnetic spin fluctuations are most pronounced, it is widely believed that the AF spin fluctuations are playing an important role in the superconducting pairing mechanism.

Nevertheless, it is still debated whether the coexistence of AF order and superconductivity in the underdoped regime occurs in all iron arsenide superconductors. A true coexistence and competition between bulk AF and superconducting orders has been established in underdoped samples of Sm-1111[11,12] and especially in single crystals of $Ba(Fe_{1-x}Co_x)_2As_2$.[13–16] Such a coexistence was not observed in La-1111[17]. Moreover, in $Ba_{1-x}K_xFe_2As_2$[18,19] and in $SmFe_{1-x}Ru_xAsO_{0.85}F_{0.15}$[20] and $Ba(Fe_{1-x}Ru_x)_2As_2$[21] it was reported to involve a macroscopically phase segregated state. Albeit, for $Ba_{1-x}K_xFe_2As_2$ a recent study established that the macroscopic phase segregation does not occur in high quality samples where the AF and superconducting orders truly coexist on the nanometer scale.[22] This suggests that the macroscopic phase segregation previously reported for $Ba_{1-x}K_xFe_2As_2$ single crystals[18,19] is of chemical origin, likely due to a variation in the K content.[23]

Even for the $Ba(Fe_{1-x}Co_x)_2As_2$ system, which is rather well investigated thanks to the availability of sizeable and fairly homogeneous single crystals, it remains to be investigated in detail how exactly the magnetic order evolves and disappears around optimum doping.

Here we present such a detailed $\mu$SR study of the dependence of the magnetic properties on Co doping in a series of $Ba(Fe_{1-x}Co_x)_2As_2$ single crystals with $0 \leq x \leq 0.15$. It shows that the magnetic order parameter is anomalously suppressed as superconductivity emerges in the underdoped regime. It also details how the static magnetic correlations disappear around optimum doping, and presents evidence that slow spin fluctuations persist in the overdoped regime. Notably, we find that a competition between the static antiferromagnetic order and superconducting orders occurs only in the underdoped regime whereas near optimum doping and even in the overdoped regime there are signatures of a cooperative relationship between inhomogeneous static or slowly fluctuating magnetic correlations and superconductivity. We remark that our results confirm previous $\mu$SR work which already reported the coexistence and competition of bulk magnetic and superconducting orders for the underdoped sample at $x$ = 0.055[15] as well as the development of an inhomogeneous magnetic state right below $T_c$ in a near optimally doped sample.[24]

## 2. Materials and Methods

The Ba(Fe$_{1-x}$Co$_x$)$_2$As$_2$ single crystals were grown from self-flux in glassy carbon crucibles. As described in Refs.[25,26], their Co content was determined by energy dispersive x-ray spectroscopy with an accuracy of about 0.05. The superconducting transition temperature, $T_c$, as shown in Fig. 1 was determined from resistivity and dc magnetization measurements that were performed using a physical property measurement system (PPMS) from Quantum Design (Model QD6000). The antiferromagnetic transition temperature, $T_N$, was deduced from the $\mu$SR experiments (as shown below). The obtained phase diagram of $T_c$ and $T_N$ as shown in Fig. 1 agrees reasonably well with previous reports.[27–31] The $x$ = 0.055 sample has been already previously investigated with $\mu$SR- and infrared spectroscopy.[15] Infrared measurements have also been performed at $x$ = 0.065.[32] Specific heat measurements on crystals from the same or similar growth batches are reported in Ref.[26]. These measurements suggest a bulk superconducting state of the samples in the Co doping range of 0.035 ≤ $x$ ≤ 0.13 with a maximum transition temperature $T_{c,max}$ = 24.5 K at $x$ = 0.065.

The muon-spin-rotation ($\mu$SR) measurements have been performed on large single crystals or on mosaics of several smaller pieces with freshly cleaved surfaces. The zero-field (ZF), longitudinal field (LF), and transverse field (TF) measurements have been conducted with the general purpose spectrometer (GPS) setup at the $\pi$M3 beamline of the Swiss Muon Source (S$\mu$S) at the Paul Scherrer Institut (PSI) in Villigen, Switzerland.

The $\mu$SR technique measures the time-resolved spin polarization of an ensemble of muons that reside on interstitial lattice sites of the studied material. A beam of fully spin-polarized so-called surface muons is produced at a proton accelerator and implanted with an average energy of about 4.2 MeV in the sample where it thermalizes rapidly and without a significant loss in spin polarization. In the iron arsenides the muons have been shown to stop at well-defined interstitial lattice sites close to an As ion.[33] The average muon implantation depth (and thus the spread of the muon stopping sites) is about 100–200 $\mu$m, the magnetic and superconducting properties probed by the muon ensemble are therefore representative of the bulk.

The muon spin polarization is recorded via the detection of the asymmetry of the emission of the positrons that arise from the radioactive decay of the muons. The obtained $\mu$SR spectra cover a time window of about $10^{-6}$ to $10^{-9}$ s. With the gyromagnetic ratio of the positive muons of $\gamma_\mu = 2\pi \times 135.5$ MHz T$^{-1}$, this enables one to detect internal magnetic fields from 0.1 G to several Tesla. The muon spins are precessing in the local magnetic field, $\mathbf{B}_\mu$, with a frequency of $\nu_\mu = \gamma_\mu \mathbf{B}_\mu / 2\pi$. The positive muons decay within an average lifetime of $\tau_\mu \approx 2.2$ $\mu$s into two neutrinos and a positron. The latter is preferentially emitted along the direction of the muon spin at the instant of decay. By detecting the time-dependence of the spatial asymmetry of the positron emission rate, the time resolved spin polarization $P(t)$ of the muon ensemble is thus obtained. The initial asymmetry is 27-28% for the measurements performed in ZF geometry and 21-22% for the ones performed in TF geometry (since the spin rotator at the GPS beamline rotates the muon spin only by 56 degree). The indicated, small differences in the initial asymmetry arise from a variation in the size and mounting of the samples as well as from the so-called veto counter that was used for the smaller samples. More details regarding the $\mu$SR technique can be found in Ref. [34–36].

The $\mu$SR technique yields the magnetic field distribution on a microscopic scale and is therefore very well suited for the investigation of the magnetic and superconducting properties of new materials. In particular, it enables a reliable determination of the volume fractions of the magnetic and superconducting phases and it can also be used to determine the temperature and doping dependence of the magnetic and superconducting order parameters. It has previously been very successfully applied to study the coexistence of magnetism and superconductivity in a variety of unconventional superconductors like the underdoped cuprates,[37–40] the ruthenate-cuprates,[41] the triplet superconductor $Sr_2RuO_4$[42] or more recently the iron arsenides[11,12,15,17,19,20,43–47] and iron selenides.[48–50]

The neutron diffraction experiments were performed at the thermal four circle single crystal diffractometer TRICS at the spallation neutron source SINQ at PSI. A pyrolytic graphite (PG-002) monochromator was used with neutron wavelength $\lambda = 2.4$ Å. The crystal was mounted in a closed cycle refrigerator and the intensities of 15 accessible magnetic Bragg peaks were measured.

Additional neutron diffraction experiments were performed at the cold neutron triple axis spectrometer RITA-II to measure selected magnetic Bragg peaks with a higher accuracy and better resolution. A wavelength $\lambda = 4.04$ Å from a PG-002 monochromator was used and the crystal was mounted in a Helium cryostat in two different orientations allowing for scans in reciprocal space along the **a** (longitudinal) and **b** (transverse) directions through the magnetic

Bragg peak at $\mathbf{Q}_{AFM} = (1,0,3)$ in orthorhombic notation. The scan directions are explained in detail in reference.[51]

### 3. Results and Discussion

#### 3.1. Magnetism in non-superconducting samples at $0 \leq x \leq 0.03$

Figure 2 shows representative low-temperature ZF-$\mu$SR spectra of the Ba(Fe$_{1-x}$Co$_x$)$_2$As$_2$ crystals that give an overview how the antiferromagnetic (AF) order evolves as a function of the Co substitution. In the undoped parent compound, at $x = 0$, it has been established by magnetic neutron diffraction that a long-range antiferromagnetic order with an Fe moment of about 0.9-1 $\mu_B$ develops below $T_N = 134$ K.[13,52–55] The corresponding ZF-$\mu$SR spectrum in Fig. 2(a) is also characteristic of a bulk, long-range ordered AF state. Its oscillatory signal has a large amplitude and a small relaxation rate similar as in previous $\mu$SR experiments.[22,56,57] The fit shown by the solid line has been obtained with the function:

$$P(t) = P(0)\left[\sum_{i=1}^{2} A_i \cdot \cos(\gamma_\mu \cdot \mathbf{B}_{\mu,i} \cdot t + \varphi) \cdot exp(-\lambda_i t) + A_3 \cdot exp(-\lambda_3 t)\right] \quad \text{eq. (1)}$$

where $P$, $A_i$, $\mathbf{B}_{\mu,i}$, $\varphi$, $\lambda_i$ are the polarization of the muon spin ensemble, the relative amplitudes of the different signals, the local magnetic field at the muon sites, the initial phase of the muon spin, and the exponential relaxation rates, respectively. The first two terms yield oscillation frequencies of about 28.4 and 7 MHz. It was previously shown for the 1111 system that these two precession frequency arise from different interstitial muon sites that are located within or between the FeAs layer, respectively.[33] By analogy we assume that two similar muon sites exist for the Ba(Fe$_{1-x}$Co$_x$)$_2$As$_2$ system. The non-oscillatory and slowly relaxing signal described by the third term arises due to the non-orthogonal orientation of the muon spin polarization, $\mathbf{P}$, and the local magnetic field $\mathbf{B}_\mu$. In polycrystalline samples with randomly oriented grains (and thus randomly oriented $\mathbf{B}_\mu$) this yields a so-called "one third tail" with $A_3 = 1/3$. For single crystals $A_3$ varies between zero and unity as the orientation between $\mathbf{B}_\mu$ and **P** changes from parallel to perpendicular. At $x = 0$ we obtain $A_3 \approx 0.3$ which suggests that the angle between $\mathbf{B}_\mu$ and **P** is about 70°. This value does not change much upon Co substitution at $0 \leq x \leq 0.055$ where the samples maintain a bulk magnetic state at low temperature that is established from TF-$\mu$SR experiments as shown below (in Fig. 8) and for the same $x = 0.055$ sample in Ref. [15].

Figures 2(b) and 2(c) show that the ZF-$\mu$SR spectra for the non-superconducting samples at $x = 0.02$ and 0.03 are only moderately affected by the Co substitution. The highest precession frequency, which is well resolved and measures the magnitude of the AF order parameter, decreases only by about 10% from 28.4 MHz at $x = 0$ (where neutron diffraction yields a magnetic moment of 0.9-1 $\mu_B$[13,52–55]) to 24.8 MHz at $x = 0.03$. The Co doping still gives rise to a sizeable increase of the relaxation rate from 3.3 $\mu s^{-1}$ at $x = 0$ to about 25 $\mu s^{-1}$ at $x = 0.03$ which signifies a significant broadening of the distribution of the local magnetic fields. These trends are shown in Fig. 3 which displays the doping dependence of the normalized value of the internal magnetic field, $\mathbf{B}_\mu(x) / \mathbf{B}_\mu(x = 0)$, and of its relative spread, $\Delta\mathbf{B}_\mu(x) / \mathbf{B}_\mu(x)$. Figure 4(a) shows for the example of the $x = 0.02$ sample how the values of $T_N$ and $\mathbf{B}_\mu$ have been deduced from the $T$-dependent ZF-$\mu$SR data.

### 3.2. Coexistence of bulk magnetism and superconductivity in underdoped samples at $0.04 \leq x \leq 0.055$

The ZF-spectra in Figures 2(d) - 2(f) and the evolution of $\mathbf{B}_\mu(x)$ and $\Delta\mathbf{B}_\mu(x)$ in Fig. 3 highlight that the suppression of the AF order parameter and its spatial variation start to evolve much more rapidly as soon as superconductivity emerges at $x \geq 0.04$. The sample at $x = 0.04$ with $T_c = 6$ K is a borderline case where our $\mu$SR data do not provide a solid proof that superconductivity is a bulk phenomenon. Nevertheless, already at $x = 0.045$ our $\mu$SR data provide firm evidence for a bulk superconducting state. Figure 5(a) shows a so-called TF-$\mu$SR pinning experiment which confirms that a strongly pinned superconducting vortex lattice develops in the entire samples volume. The sample was initially cooled in an external field of $\mathbf{H}^{ext}$ = 500 Oe to $T$ = 1.6 K $\ll T_c$ = 13 K. The first TF-$\mu$SR measurement was then performed under regular field cooled conditions at $T$ = 1.6 K. The so-called $\mu$SR lineshape, as obtained from a fast-Fourier transformation of the TF-$\mu$SR time spectrum, is shown by the solid symbols. It details the distribution of the local magnetic field that is probed by the muon ensemble. This lineshape consists of a very narrow peak due to about 2-3% of background muons that stop outside the sample, in the sample holder or the cryostat walls and windows. The muons stopping inside the sample give rise to the very broad main peak whose unusually large width is caused by the static magnetic order which persists throughout the entire sample volume. The broadening due to the superconducting vortex lattice is in comparison much smaller and completely overwhelmed by the magnetic contribution. The magnetic order is documented by the ZF-$\mu$SR spectrum in Fig. 5(b) which reveals a strongly damped, oscillatory signal with an average frequency of 20.4 MHz. Before the second TF-$\mu$SR measurement (open symbols) the external magnetic field was increased by 250 Oe to $\mathbf{H}^{ext}$ = 750 Oe while the temperature was kept at 1.6 K $\ll T_c$. It is evident from Fig. 5(a) that only the narrow peak due to the background muons follows the change of $\mathbf{H}^{ext}$ whereas the broad peak due to the muon inside the sample remains almost unchanged. This observation, that the magnetic flux density inside the sample remains unchanged, is the hallmark of a bulk type-II superconductor with a strongly pinned vortex lattice. Notably, this bulk superconducting vortex state coexists with an antiferromagnetic order that is also a bulk phenomenon. The combined ZF-$\mu$SR and TF-$\mu$SR data thus provide unambiguous evidence that the superconducting and magnetic orders coexist on a nanometer scale. A corresponding pinning effect due to a superconducting vortex lattice that exists in the presence of a bulk magnetic order has also been observed at $x = 0.05$ (not shown) and it was previously demonstrated for a crystal at $x = 0.055$.[15]

Clear signatures of the competition between the magnetic and superconducting orders have already been reported from neutron diffraction[13,14] and later from TF-$\mu$SR measurements.[15] The combined neutron and $\mu$SR data revealed that the magnitude of the magnetic order parameter is anomalously suppressed below $T_c$ while the volume fraction of the magnetic phase remains close to 100%.[15] In Fig. 6 we compare a set of magnetic neutron diffraction and $\mu$SR data that have been obtained on the very same single crystal with $x = 0.05$. Transverse and longitudinal scans through the magnetic Bragg peak at $\mathbf{Q}_{AFM} = (1,0,3)$ in orthorhombic notation are shown in Fig. 6 (a) and 6 (b), respectively. Single resolution limited peaks are observed in both scan directions, indicating commensurate antiferromagnetic order with a correlation length in excess of 300 Å. The diffraction data obtained at $T = 5$ K are consistent with a stripe type C- antiferromagnetic structure with an ordered Co moment of ~0.1 $\mu_B$ per Fe ion. The moment is thus reduced by about a factor of 10 from the value in the parent compound $BaFe_2As_2$.[55] This agrees well with the value deduced from the $\mu$SR relaxation rate as shown in Fig. 3. Figure 6(c) compares the $T$-dependence of the peak intensity (solid symbols) with the one of the TF-$\mu$SR relaxation rate $\lambda^{TF}$ (open symbols). The neutron and $\mu$SR data consistently reveal the onset of the static magnetic order below $T_N \approx 40$ K. They both also exhibit a pronounced anomaly around $T_c \approx 19$ K which signifies the superconductivity-induced suppression of the magnetic order parameter.

The direct comparison of the neutron diffraction and $\mu$SR data also yields important information with respect to the magnetic order. The observation of well-defined, resolution limited Bragg-peaks in magnetic neutron diffraction, as shown in Figs. 6(a) and (b), demonstrates that the AF order is commensurate and fairly long-ranged. It is not in agreement with the incommensurate AF order which has been reported from NMR experiments.[58,59] The incommensurability of $\varepsilon \approx 0.04$ as reported in Ref. [59] would be well beyond the resolution limit of the neutron diffraction experiment presented in this paper and is therefore clearly not observed. On the other hand, the rapid exponential relaxation of the ZF-$\mu$SR signal in Fig. 6(d), with no trace of an oscillation, implies that the spread of the local magnetic field must be fairly large. The neutron and $\mu$SR techniques appear to be probing the same kind of magnetic order since they yield the same onset temperature $T_N = 40$ K, the same anomalous suppression around $T_c$, and very similar values of the order moment of ~0.09-0.1 $\mu_B$ per Fe ion. A likely explanation therefore is in terms of a sizeable spatial variation of the amplitude of the commensurate AF order. This variation could be induced by the randomly distributed Co atoms which give rise to a spatial variation of the magnitude of the antiferromagnetic order parameter. We note that a similar conclusion was obtained from previous NMR experiments on Ni substituted Ba-122.[60]

Further information about the role of the Co-induced disorder in the suppression of the magnetism can be obtained from the comparison with the $\mu$SR data on the hole doped Ba$_{1-x}$K$_x$Fe$_2$As$_2$.[22] The substitution-induced disordering effects should be significantly weaker here since the K ions are incorporated on the Ba sites where they do not directly disturb the iron arsenide layers. The relaxation rate of the oscillatory signal in the ZF-$\mu$SR spectrum of the Ba$_{1-x}$K$_x$Fe$_2$As$_2$ sample with $x = 0.19$ in Ref. [22] is indeed significantly smaller than the one of the Ba(Fe$_{1-x}$Co$_x$)$_2$As$_2$ crystal at $x = 0.045$ which is in a similarly underdoped state with $T_c \approx 0.5T_{c,max}$. Nevertheless, in both the K- and the Co substituted samples the $\mu$SR precession frequency and thus the magnetic order parameter are only moderately reduced as long as the samples are not yet superconducting or remain strongly underdoped with $T_c \leq 0.5 T_{c,max}$. This common behavior suggests that the disorder effects in the Co substituted samples are not governing the general features of the magnetic and superconducting phase diagram. Figure 3 also shows that in Ba(Fe$_{1-x}$Co$_x$)$_2$As$_2$ the magnetic order parameter is very rapidly suppressed between $x = 0.045$ and $x = 0.05$ where the Co concentration increases by a relatively small amount. It was previously shown that the weak magnetic state at $x = 0.055$ remains a bulk phenomenon[15] with a well-defined commensurate order.[13,14] These observations rather point toward an intrinsic origin of the transition, for example due to a change of the Fermi-surface topology as it was observed in this doping range with the angle resolved photo-emission spectroscopy (ARPES).[61] A corresponding $\mu$SR study of homogeneous Ba$_{1-x}$K$_x$Fe$_2$As$_2$ samples in the weakly underdoped regime with $0.5 T_{c,max} < T_c < T_{c,max}$, to the best of our knowledge is still lacking.

Another interesting feature concerns the very different doping dependences of $T_N$ and $\mathbf{B}_\mu$. Figure 3 displays the monotonous and almost linear suppression of $T_N$ in the range $0 \leq x \leq 0.055$ with no sign of an anomaly between $x = 0.045$ and $0.05$ where $\mathbf{B}_\mu$ is suddenly reduced from ~70% to ~10% of its value at $x = 0$. Such a qualitatively different behavior of $T_N$ and $\mathbf{B}_\mu$ is rather surprising since the thermodynamic properties such as the transition temperature, $T_N$, should be governed by the magnetic order parameter. It implies that the actual value of $T_N$ is much lower than the upper bound that is set by the magnetic order parameter. It has indeed been found that $T_N$ is closely linked to the structural transition temperature, $T_s$, where the crystal symmetry changes from tetragonal to orthorhombic. The $T_N$ values are always slightly below or at best equal to $T_s$ which suggests that the orthorhombic distortion is a prerequisite for the static magnetic order to develop.[31,62,63] It is still debated whether the orthorhombic transition itself is caused by an electronic instability. The proposed explanations range from the scenario of an orbital polarization due to a different occupation of the Fe-$d_{xz}$ and $d_{yz}$ orbitals[64,65] to a nematic instability of the electronic system that is driven by the anisotropic magnetic fluctuations.[63] In the presence of a sizeable magneto-elastic coupling,[31] the former model naturally explains that $T_N$ is tied to $T_s$ and thus may not be strongly affected by the rapid suppression of the magnetic order parameter. In the context of the nematic model such a different behavior of $T_N$ and $\mathbf{B}_\mu$ is less obvious since the magnetic correlations are at the heart of both the structural and the magnetic transitions. As outlined in Ref. [63], the hierarchy of the nematic and the magnetic transition is determined by their different sensitivity to fluctuation effects. The nematic transition is less sensitive since it breaks only a discrete lattice symmetry whereas the AF transition requires in addition that the continuous rotational symmetry is broken. While this may explain that the anomaly in $T_N$ is considerably weaker than the one in $\mathbf{B}_\mu$, it still makes it difficult to understand that this anomaly is essentially absent.

### 3.3. Spatially inhomogeneous magnetic order around optimum doping at $0.055 < x < 0.075$

Next, we address the question how the static magnetic order vanishes around optimal doping. We show that this involves a spatially inhomogeneous magnetic state for which the magnetic volume fraction decreases systematically with increasing Co content and finally vanishes at $x \geq 0.075$. Notably, we find that this inhomogeneous magnetic state develops right below $T_c$ which suggests that it may have a constructive rather than a competitive relationship with superconductivity. The evolution of the magnetic properties in the range from $x = 0.055$ to 0.075 is captured in Fig. 7 which displays the temperature dependence of the ZF-$\mu$SR spectra and in Fig. 8 which shows the corresponding low temperature TF-$\mu$SR spectra. The ZF-$\mu$SR spectra at $T > T_N$ have been fitted with a so-called Kubo-Toyabe function which describes the weak depolarization due to the nuclear moments. This function has been multiplied with an exponential function with a relaxation rate $\lambda^{ZF}$ to account for an additional relaxation due to slow magnetic fluctuations. The spectra at $T < T_N$ were fitted with a sum of two exponential functions one of which has a large relaxation rate to describe the rapidly depolarizing part of the signal. The amplitude of this rapidly relaxing signal is shown in Figs. 7(e) - 7(h) as a function of doping and temperature. The obtained normalized values of $T_N$ and $\mathbf{B}_\mu$ in the spatially inhomogeneous phase are shown by the open symbols in Fig. 3.

In the bulk magnetic state at $x = 0.055$, the rapidly depolarizing ZF-$\mu$SR signal develops below $T_N \approx 32$ K and its amplitude reaches about 65% as shown in Figs. 7(a) and 7(e). As discussed above, the slowly relaxing part of the ZF-$\mu$SR signal of about 35% arises in these single crystals because the muon spin polarization and $\mathbf{B}_\mu$ are not orthogonal. This is confirmed by the TF-$\mu$SR data in Fig. 8 where at $x = 0.055$ the entire TF-$\mu$SR signal depolarizes very rapidly (except for a small fraction due to the background muons that stop outside the sample). Figures 7(b) and 7(c) show that the ZF-$\mu$SR spectra at $x = 0.06$ and $0.065$ exhibit a noticeably different behavior since the amplitude of the magnetic signal is significantly reduced here. This trend is also evident from the TF-$\mu$SR data in Fig. 8 where the amplitude of the rapidly depolarizing signal also decreases systematically. The solid lines show a two component fit with a rapidly depolarizing component that describes the magnetic regions and a more slowly depolarizing one that accounts for the nonmagnetic regions. In the latter the relaxation is dominated by the superconducting vortex lattice which apparently develops in both the nonmagnetic and the magnetic regions as is evident from additional pinning experiments (not shown). From the fits of the TF-$\mu$SR data we deduced that the volume fraction of the magnetic phase decreases from essentially 100% at $x = 0.055$ to ~50% at $x = 0.06$ and ~30% at $x = 0.065$.

Notably, we find that this spatially inhomogeneous magnetic order develops only in the superconducting state. Figure 7(f) and 7(g) show indeed that the onset of the magnetic signal coincides with $T_c$. This curious coincidence was observed in several other near optimally doped crystals with $T_c$ values close to $T_{c,max}$ = 25 K, an early example is reported in Ref. [24]. Some of these crystals may have contained impurity phases and their Co content may not always have been well controlled and characterized, they still establish the general trend that the onset of this inhomogeneous magnetic order coincides with $T_c$. Figures 7(d) and 7(h) show that this inhomogeneous magnetic order is finally absent in the crystal with $x$ = 0.075 that is moderately overdoped with $T_c \approx$ 21.7 K. It is also not observed in the more strongly overdoped crystals at $x$ = 0.09, 0.11 and 0.12.

The systematic decrease of the magnetic volume fraction within the fairly narrow window of $0.055 \leq x \leq 0.075$ around optimum doping makes it seem unlikely that a magnetic impurity phase is responsible for the spatially inhomogeneous magnetic order. This could also not explain the observation that the onset of the magnetic order coincides with $T_c$.

The same arguments apply against the scenario that the magnetic order may be a muon-induced effect, either due to its positive charge or a lattice distortion. There is no obvious reason why such an effect should only occur in the superconducting state and if so only in parts of the sample volume. Besides, we note that at $x$ = 0.05 we obtained a good agreement between the neutron diffraction and the $\mu$SR data (see Fig. 4) which confirms that the muons do not noticeably disturb the magnetic order. This is despite the rapid change of the magnetic properties in the range of $0.045 < x < 0.055$ which should enhance the susceptibility to any muon-induced effects.

The magnetic neutron scattering experiments revealed a change of the magnetic ground state between $x$ = 0.055 and 0.06. The AF order is reported here to become incommensurate[51] and eventually to exhibit a reentrant behavior.[31] A similar behavior was recently reported for Ni substituted crystals where also a sudden transition from a commensurate to an incommensurate AF order was observed.[66] The latter was still found to develop at $T_N > T_c$ and to compete with superconductivity. From Fig. 7 it can be seen that our $\mu$SR data at $0.055 < x$

< 0.075 show no evidence of such a reentrant behavior where the magnetic order is weakened or even vanishes at low temperature. Instead, at $x = 0.06$ and 0.065 we find that the magnetic order develops right below $T_c$ with the amplitude of the magnetic signal increasing toward low temperature. A possible explanation for these different trends may be that the spatially inhomogeneous magnetic state seen by $\mu$SR near optimal doping is strongly disordered and yields magnetic Bragg peaks that are very broad and difficult to observe in neutron diffraction. This broadening may be enhanced in the superconducting state such that the reentrant behavior reported in near optimally doped samples in Ref. [31] may actually arise from a broadening of the magnetic Bragg peaks. Clearly, further efforts should be undertaken to investigate this inhomogeneous magnetic state at $0.055 < x < 0.075$ with magnetic neutron diffraction.

### 3.4. Superconductivity-induced enhancement of spin fluctuations in overdoped samples

Finally, we show that an anomalous, superconductivity-induced enhancement of the low energy spin fluctuations occurs even in strongly overdoped samples. Figure 9 shows the ZF- and LF-$\mu$SR data for an overdoped crystal with $x = 0.11$ and $T_c = 10$ K. From the ZF-$\mu$SR spectra in Fig. 9(a) it can already be seen that the relaxation rate in the superconducting state is slightly larger than the one in the normal state. The effect is relatively weak since the relaxation in these ZF spectra is dominated by the contribution of the randomly oriented nuclear magnetic moments which is described by a so-called Kubo-Toyabe function. It is well known that a small longitudinal field (LF) can be used to reduce this static nuclear contribution as to reveal the weaker dynamical relaxation due to low energy spin fluctuations.[36] The corresponding LF-$\mu$SR data for small longitudinal fields of 5 Oe and 10 Oe are shown in Figs. 9(b) and 9(c), respectively. They exhibit small yet clearly visible changes that are characteristic of an increase of the dynamical relaxation rate at low temperature. Figure 9(d) details the temperature dependence of the dynamical relaxation rate, $\lambda^{LF}$, for LF = 10 Oe which exhibits a clear anomaly at $T_c \approx 10$ K. The sudden increase of $\lambda^{LF}$ below $T_c$ signifies a superconductivity-induced enhancement of the low energy spin fluctuations in this strongly overdoped crystal. The inset of Fig. 9(d) shows how this SC-induced increase in relaxation rate, $\Delta\lambda^{LF}$, evolves as a function of the longitudinal field. The green line shows a fit with the so-called Redfield-function

$$\Delta\lambda^{TF} = 2(\gamma\mu_o H_\mu)^2 \tau_c / [1 + (\gamma\mu_o H^{LF}\tau_c)^2] \qquad \text{eq. (2)}$$

which describes a relaxation process due to fluctuating local fields[36] with a magnitude, $\mu_o H_\mu =$ 0.16 G , and a correlation time, $\tau_c = 0.65$ $\mu$s.

In this context we recall a related trend in the TF-$\mu$SR data where it was noticed that in the superconducting state the $\mu$SR lineshape exhibits an anomalous paramagnetic shift.[46,48,67] This superconductivity-induced paramagnetic shift of the TF-$\mu$SR lineshape, which has also been consistently observed in the present optimally and overdoped single crystals (not shown), is most likely related to the superconductivity-induced enhancement of the low energy spin fluctuations.

The $\mu$SR data thus provide compelling evidence that even in the optimally doped and overdoped samples, the magnetic correlations and/or the low-energy spin fluctuations are noticeably enhanced in the superconducting state. A similar behavior was previously observed in materials with a spin-triplet superconducting order parameter that breaks time reversal symmetry, like in $Sr_2RuO_4$[42], $PrOs_4Sb_{12}$[68] or $LaNiC_2$[69]. However, such a spin-triplet state does not seem to be realized in $Ba(Fe_{1-x}Co_x)_2As_2$ for which NMR Knight shift measurements revealed a pronounced superconductivity-induced reduction of the Knight shift that is a hallmark of a spin-singlet superconducting state.[70,71] Therefore, it appears that alternative explanations for the enhanced spin correlations in the superconducting state of optimally doped and overdoped Ba(Fe$_{1-x}$Co$_x$)$_2$As$_2$ need to be explored. This issue is beyond the scope of our present $\mu$SR study.

## 4. Summary

With muon spin rotation ($\mu$SR) we investigated the magnetic and superconducting phase diagram of a series of Ba(Fe$_{1-x}$Co$_x$)$_2$As$_2$ single crystals with $0 \leq x \leq 0.15$. We showed that the magnetic order parameter is initially only weakly reduced at $x \leq 0.04$, whereas it exhibits a much more rapid suppression as superconductivity emerges at $x > 0.04$. In most of the underdoped regime we observed a weakened, yet bulk magnetic order that coexists and competes with superconductivity. The comparison with neutron diffraction data suggests that this AF order remains commensurate. The Co atoms likely induce a random variation of the amplitude which leads to a large $\mu$SR depolarization rate. The final suppression of the magnetic order toward optimum doping involves a spatially inhomogeneous magnetic state. The magnetic volume fraction becomes rapidly reduced at $x > 0.055$ and it vanishes at $x = 0.075$. Notably, this inhomogeneous magnetic order develops right below $T_c$ and therefore does not seem to compete with superconductivity but rather seems to have a constructive relationship. Even in a strongly overdoped samples at $x = 0.11$ the $\mu$SR experiments reveal signatures of a superconductivity-induced enhancement of the low-energy spin fluctuations. Our observations highlight a versatile relationship between magnetism and superconductivity that is competitive on the underdoped side of the Co-doping phase diagram whereas it appears to be cooperative in optimally doped and overdoped samples.

**Acknowledgement**

We acknowledge fruitful discussions with Dr. Christoph Meingast. The $\mu$SR experiments were performed at the Swiss muon Source (S$\mu$S) at the Paul Scherrer Institut, Villigen,

Switzerland where we benefitted from the technical support of Dr. Alex Amato and Dr. Hubertus Lütkens. We appreciate Bente Lebech for helping us with the refinement of the magnetic neutron diffraction data to obtain absolute values of the magnetic moment. The work at University of Fribourg has been supported by the Swiss National Science Foundation (SNF) grants 200020-129484 and 200020_140225, by the NCCR MaNEP and the project no. 122935 of the Indo-Swiss Joint Research Program (ISJRP). The work in QMUL was supported by the Leverhulme Trust and C.A. was funded by an EPSRC Doctoral Training Award. The work in Karlsruhe was supported by the Deutsche Forschungsgemeinschaft (DFG) through SPP 1458. The financial support from D.S.T. (Government of India) and M.H.R.D. (Government of India) is highly acknowledged.

*Present Address: Microelectronics Research Center, University of Texas at Austin, Texas 78758, USA

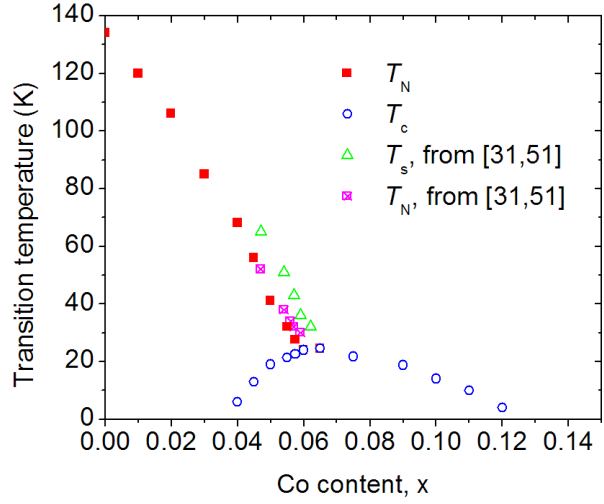

**Figure 1:** Phase diagram of the magnetic transition temperature, $T_N$, as determined with $\mu$SR and the superconducting critical temperature, $T_c$, obtained from resistivity and magnetic susceptibility measurements as well as from the specific heat data of Ref. [26] shown for comparison are the magnetic and structural transition temperatures of Refs. [31,51].

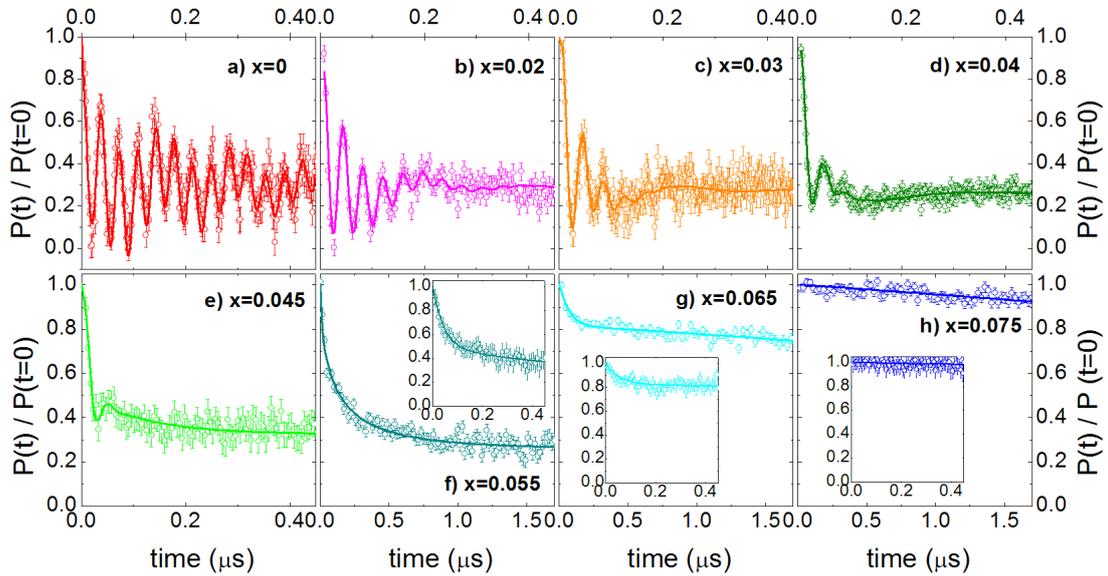

**Figure 2:** Low temperate ZF-$\mu$SR spectra of Ba(Fe$_{1-x}$Co$_x$)$_2$As$_2$ single crystals with $0 \leq x \leq 0.075$ showing how the magnetic order is suppressed by the Co substitution. Insets in (f) – (h) show the spectra at early times as in (a) – (e).

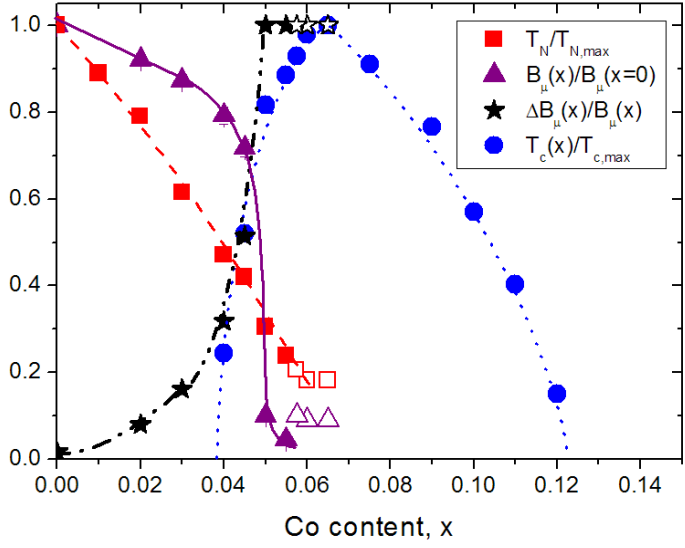

**Figure 3:** Phase diagram showing the Co-dependence of the normalized magnetic and superconducting transition temperatures, $T_N$ and $T_c$, and the normalized values of the average magnetic field at the muon site, $\mathbf{B}_\mu$, and of its relative spread, $\Delta \mathbf{B}_\mu$. The open symbols show the magnetic properties in the spatially inhomogeneous magnetic state near optimum doping. For samples $0 \leq x \leq 0.045$ with an oscillatory signal in the ZF-$\mu$SR time spectra $\mathbf{B}_\mu$ and $\Delta \mathbf{B}_\mu$ have been obtained by fitting with the function in equation (1). For samples $0.05 \leq x \leq 0.065$ where the oscillatory signal is completely overdamped (see Fig. 2) only exponential relaxation functions were used and $\mathbf{B}_\mu$ was estimated from the relaxation rate according to $\lambda^{ZF} = \gamma\, \mathbf{B}_\mu$. The latter estimate agrees with the one obtained from the TF-$\mu$SR data as shown in Fig. 4(b) for $x = 0.05$.

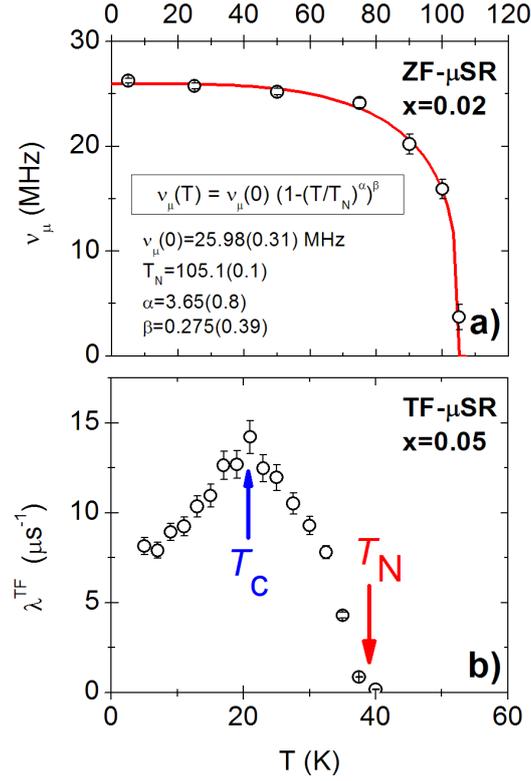

**Figure 4:** (a) $T$-dependence of the highest ZF-$\mu$SR precession frequency, $\nu_\mu$, at $x = 0.02$. The fit (solid line) shows the determination of the low temperature value of the local magnetic field, $\mathbf{B}_\mu(T = 0) = 2\pi/\gamma \cdot \nu_\mu(T = 0)$ with $\gamma = 851.4$ MHz T$^{-1}$, and the AF transition temperature, $T_N$. (b) $T$-dependence of the transverse-field ($\mathbf{H}^{ext} = 3$ kOe) relaxation rate, $\lambda^{TF}$, at $x = 0.05$. The sharp increase of $\lambda^{TF}$ marks $T_N$, the anomaly at $T_c$, below which $\lambda^{TF}$ decreases again, arises due to the competition between the magnetic and superconducting orders. From the value of $\lambda^{TF}(T_c)$ and the relationship $\lambda^{TF} = \gamma/\sqrt{2}\,\mathbf{B}_\mu$ we obtained an independent estimate of $\mathbf{B}_\mu$ that agrees with the one obtained from the ZF-$\mu$SR data as shown in Fig. 3.

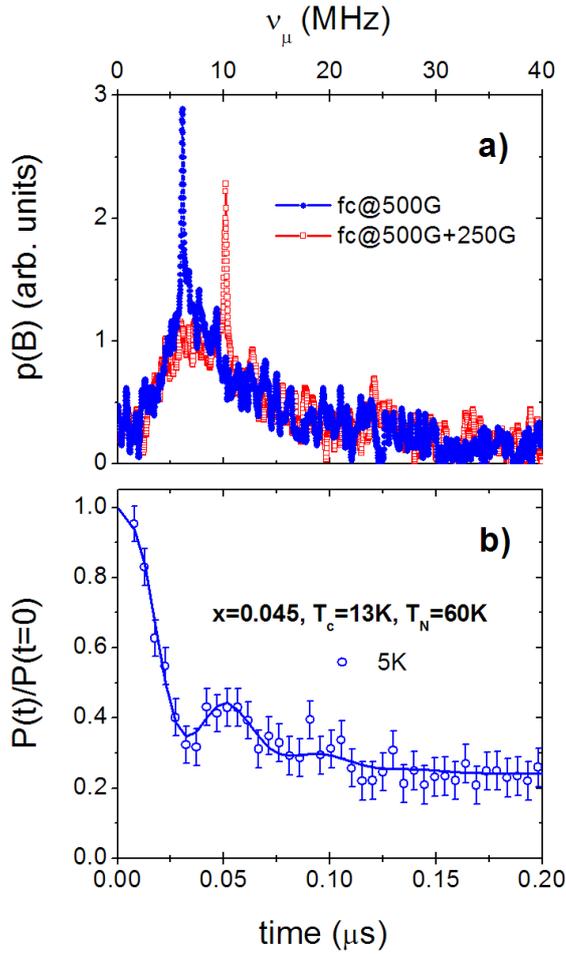

**Figure 5: (a)** TF-$\mu$SR lineshapes at $x = 0.045$ obtained during a so-called "pinning experiment" showing the presence of a strongly pinned, bulk superconducting vortex lattice. The sample was initially cooled to $T = 1.6$ K $\ll T_c = 13$ K in a transverse field of $\mathbf{H}^{ext} = 500$ Oe. The first TF-$\mu$SR lineshape (blue symbols) was measured under standard field-cooled conditions. Before the second TF-$\mu$SR lineshape measurement (red symbols), the field was increased to 750 Oe at 1.6 K. The pinning of a bulk vortex lattice is evident since only the narrow peak, due the background muons stopping outside the sample, follows the change of $\mathbf{H}^{ext}$ whereas the broader part of the $\mu$SR-lineshape, due to the muons that stop inside the sample, remains almost unchanged. **(b)** Corresponding ZF-$\mu$SR spectrum at 5 K showing a large oscillatory signal that is characteristic of a bulk magnetic order.

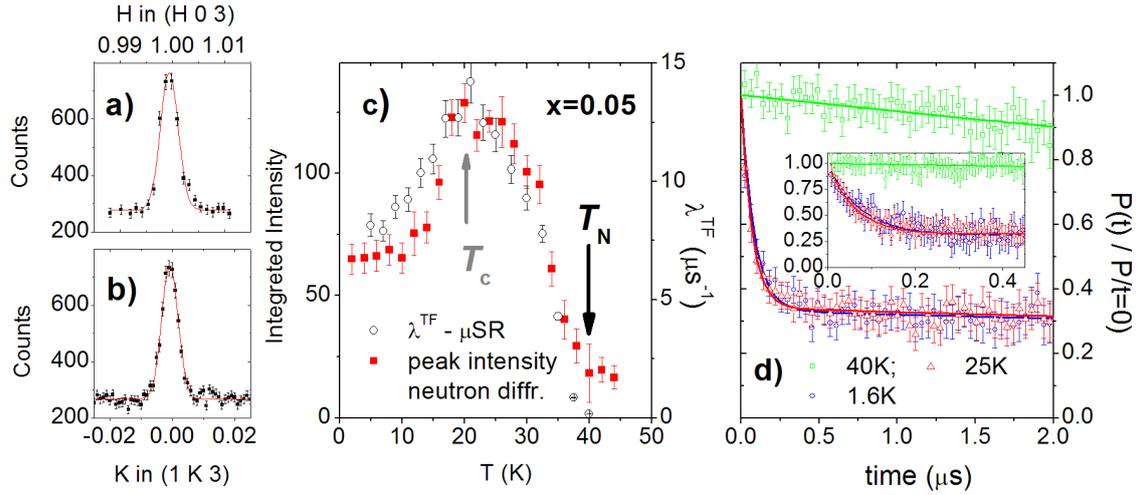

**Figure 6 :** (a) and (b) Neutron diffraction data at $x = 0.05$ showing the antiferromagnetic Bragg peak at the commensurate (1,0,3) position (in orthorhombic notation). (c) Comparison of the $T$-dependence of the intensity of the Bragg peak from neutron diffraction and the relaxation rate, $\lambda^{TF}$, of the TF-$\mu$SR experiment. The good agreement confirms that they both probe the same bulk magnetic order parameter that is suppressed below $T_c$ since it competes with superconductivity. (d) ZF-$\mu$SR spectra at $x = 0.05$ showing a rapid depolarization without any trace of an oscillatory signal. Inset: Magnification of the fast relaxation at early times.

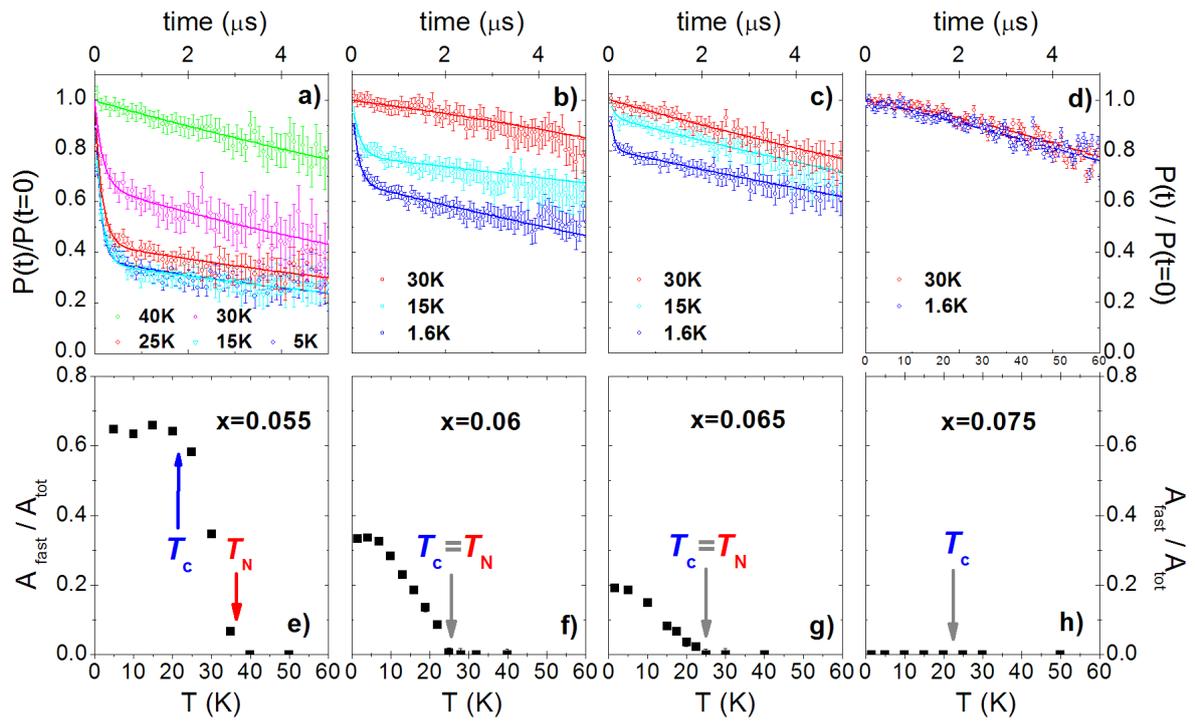

**Figure 7:** ZF-$\mu$SR spectra in the range of $0.055 \leq x \leq 0.075$ showing that the final suppression of the magnetic order around optimum doping involves a spatially inhomogeneous state with a reduced magnetic volume fraction.

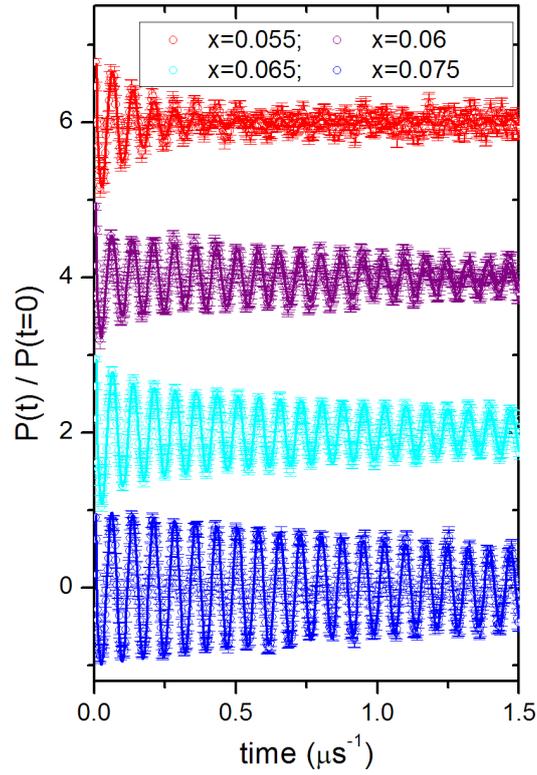

**Figure 8:** TF-$\mu$SR spectra of the nearly optimally doped crystals at $T = 1.6$ K $\ll T_c$. The spectra are shifted up by +6 for $x = 0.055$, +4 for $x = 0.06$, and +2 for $x = 0.065$. The spectra were obtained at $\mathbf{H}^{ext} = 3$ kOe and are shown in a rotating reference frame corresponding to $\mathbf{H}^{rot} = 2$ kOe.

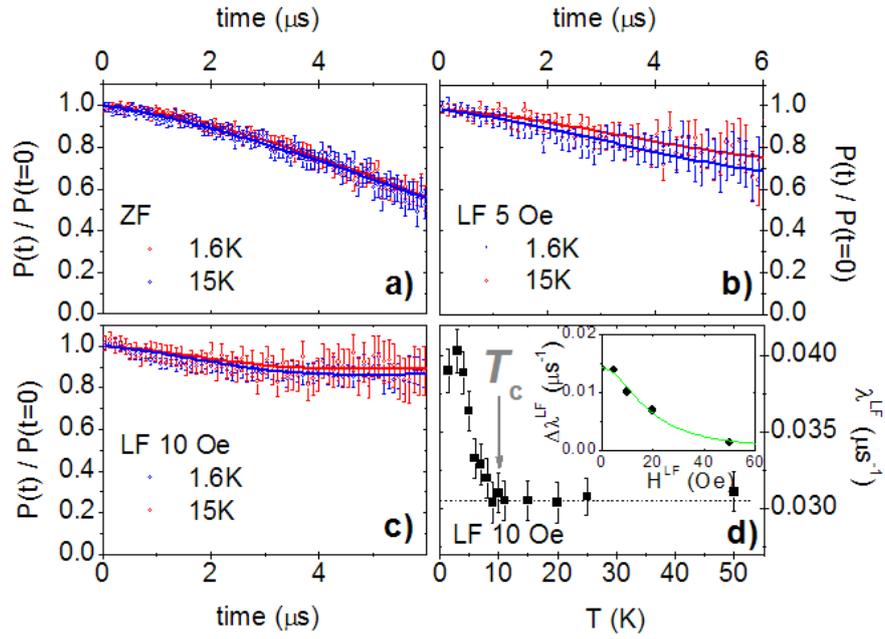

**Figure 9: (a)** to **(c)** ZF-$\mu$SR and weak LF-$\mu$SR spectra for the strongly overdoped crystal with $x = 0.11$ and $T_c \approx 10$ K. (d) Temperature dependence of the LF-$\mu$SR relaxation rate $\mathbf{H}^{LF} = 10$ Oe showing a sudden increase below $T_c$ which reveals a superconductivity-induced enhancement of the low-energy spin fluctuations. Inset: Superconductivity-induced enhancement of the relaxation rate, $\Delta\lambda^{LF}$, as a function of the longitudinal field. The green line shows a fit with the so-called Redfield-function as shown in eq.(2) and discussed in the text.